# Ultrafast probing of isotope-induced explicit symmetry breaking in ethylene

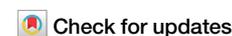 Check for updates

Alessandro Nicola Nardi ®[1,3], Alexie Boyer ®[2,3], Yaowei Hu[2], Vincent Loriot ®[2], Franck Lépine ®[2], Morgane Vacher ®[1] ✉ & Saikat Nandi ®[2] ✉

Symmetry governs nature's laws, yet many of the natural phenomena occur due to the breakdown of symmetry. Here, we show how isotope-induced inversion symmetry breaking influences ultrafast photoisomerization processes in ethylene. Using extreme ultraviolet pump – near infrared probe time-of-flight mass spectrometry, we find that replacing one of the carbon atoms in ethylene with a $^{13}$C isotope leads to twice-faster structural relaxation via ethylene-ethylidene isomerization in the photo-excited molecular cation. Advanced trajectory surface hopping calculations incorporating the nuclear symmetry of the molecular systems, reveal that it arises from the mixing of different normal modes in the isotope-substituted species, interactions otherwise forbidden by symmetry. Although the mixing does not alter the symmetry of the electronic Hamiltonian, it modifies that of the nuclear Hamiltonian, causing explicit symmetry breaking. This facilitates efficient intra-molecular vibrational energy redistribution, lowering the isomerization yield. Our findings offer opportunities to use isotope-induced nuclear symmetry breaking to control the outcome of light-molecule interactions across ultrafast timescales.

Since Pierre Curie's seminal work[1], symmetry and symmetry breaking have been known to play a central role in contemporary physics. Notable instances include gauge-symmetry breaking in superconductors[2], or, in the Higgs mechanism[3]. In these situations, the breakdown of symmetry takes place in systems that have infinite degrees of freedom, and thus an asymmetric ground state[4]. For small molecules, such as diatomic, or triatomic systems, the ground state is highly symmetric due to the reduced number of electronic and nuclear degrees of freedom. In a pioneering experiment with $H_2$ molecules, it was shown that such a symmetry can be broken by absorption of a linearly polarized photon[5]. With the development of tabletop high-order harmonic generation (HHG) based sources as well as free-electron lasers (FELs) providing attosecond (1 as = $10^{-18}$ s) and femtosecond (1 fs = $10^{-15}$ s) pulses in the short-wavelength region (few tens of nanometers to few nanometers), there is a growing interest in the ultrafast breakdown of symmetry in polyatomic molecules with high-symmetry point groups[6,7]. For instance, femtosecond transient X-ray absorption has been used to study symmetry breaking in tetrahedral molecules, such as methane[8] and tetrafluoromethane[9]. The signature of symmetry breaking in Jahn-Teller-distorted methane cation was also visualized using ultrafast Coulomb explosion imaging[10]. Attosecond charge migration in hydrogen and water molecules has been predicted to break electronic symmetry while preserving the nuclear one[11]. Carbon dioxide dimers photoionized by extreme ultraviolet (XUV) pulses from an FEL are shown to undergo femtosecond symmetry-breaking dynamics leading to charge asymmetry[12]. These studies collectively demonstrate how ultrashort radiation can induce symmetry-breaking dynamics in various molecular systems.

Here, we investigate a different approach to symmetry breaking: substituting one of the carbon atoms in a neutral ethylene molecule with a $^{13}$C isotope. This substitution reduces the molecular symmetry from $D_{2h}$ to $C_{2v}$, allowing us to explore how such a change influences light-induced relaxation dynamics across ultrafast timescales. The effect of isotope-induced inversion symmetry breaking is well documented in the frequency domain, especially for hydrogen[13–15] and nitrogen[16] molecules. Isotopic substitution changes the zero-point energy of the molecule resulting in kinetic isotope effects, which were recently investigated across few-femtosecond[17] and attosecond[18,19] timescales. The present study examines how the explicit breakdown of the inversion symmetry that preserves the symmetry of the electronic Hamiltonian, but not of the nuclear Hamiltonian, affects photo-isomerization in ethylene cation. The electronic Hamiltonian depends only on the positions and charges of the nuclei, making it insensitive to mass differences. However, the nuclear Hamiltonian includes a kinetic energy term that depends on nuclear masses, so substituting one carbon with a heavier isotope explicitly breaks the inversion symmetry in the nuclear motion. Following photoionization by a broadband XUV attosecond pulse train produced via HHG, the resulting $C_2H_4^+$ was promoted to an electronic excited state from which it could relax via different pathways[17,20]: H-loss, $H_2$-loss, ethylene–ethylidene ($CH_3CH^+$) isomerization. A time-delayed femtosecond near-infrared (NIR) pulse probed the relaxation dynamics in the cation by further ionizing and/or, dissociating it. The $CH_3^+$ fragment carried the signature of the relaxation pathway via ethylene–ethylidene conversion[20–22]. We interpreted the corresponding ion yields, measured as a function of the pump-probe delay, using non-adiabatic trajectory surface hopping simulations. These simulations revealed that the difference in relaxation timescales between the isotopologues arises from the mixing of different normal modes, otherwise symmetry-forbidden, enabled by the explicit symmetry breaking. Furthermore, our calculations showed that the isotopic substitution, accompanied by a reduction in molecular symmetry, significantly suppresses the isomerization yield.

## Results

Absorption of a photon from the attosecond pulse train (energies between 17.0 and 23.3 eV) can produce ethylene cation mainly in four different cationic states from $D_0$ to $D_3$, with only a modest contribution from the $D_4$ state (see Supplementary Note 1 for details about the energetics of the photoionization process from our ab initio calculations and previous literature). Typical two-color ion-yields as a function of the XUV pump – NIR probe delay for $^{12}CH_3^+$ from $^{12}C_2H_4$, $^{13}CH_3^+$ from $^{13}C_2H_4$ and $^{13}CH_3^+$ from $^{13}C^{12}CH_4$, are shown in Fig. 1a–c. The appearance energy for this fragment was measured to be around 20 eV, with the probability of production reaching a maximum at around 27 eV[23]. The interpretation of exponential decays observed in ion-yields in XUV pump – NIR probe experiments has been addressed in multiple studies over the years[24,25]. Here, to understand the time-dependent signal, we make the distinction between the time it takes to form the ethylidene isomer and the timescale of the dynamics via that relaxation channel[17]. The molecule in an excited cationic state relaxes non-adiabatically via certain pathways and the NIR depopulates the state by promoting a part of the population to a higher-lying dissociative state leading to the production of $CH_3^+$ ions. If the probe arrives at a delay by which time the molecule has already relaxed electronically and/or, structurally via change in nuclear coordinates, the NIR pulse may no longer promote the system to the dissociative state to generate any $CH_3^+$ ion. This causes an exponentially decaying signal as a function of the delay between the two pulses. The same mechanism has been put forward previously for describing pump-probe delay-dependent fragmentation yields in ethylene and deuterated ethylene[17]. Recently, Vismarra et al.[26] have developed a similar explanation behind observed time-dependent yields in XUV-ionized donor-π-acceptor systems. In their case, as the nuclear wave-packet evolves, the energy separation between the populated states and the dissociative final state increases, making multi-photon absorption less efficient over time. It progressively reduces the likelihood of promoting the wave-packet to the dissociative state of interest, ultimately leading to the observed decay in ion-yield. This is also conceptually consistent with the way the time-dependent signal is interpreted in the present manuscript. The corresponding experimental time constants ($\tau$) from three different targets (see Methods for details about the data analysis) are shown in Fig. 1(d). The average decay constant for $^{13}CH_3^+$ from $^{13}C^{12}CH_4$ is 50 ± 8 fs, almost twice smaller than that of 108 ± 4 fs for $^{12}CH_3^+$ from $^{12}C_2H_4$. For $^{13}CH_3^+$ from $^{13}C_2H_4$, it is 83 ± 4 fs. They do not change significantly with the NIR probe intensity, confirming that the observed time-dependent ion yield is representative of the dynamics initiated by the XUV-pump only. It is interesting to note that the observed trend in the $\tau$-values for different targets cannot be explained by mass effects[27]. To better understand the trend observed in the measurements, and to identify the relevant electronic and/or nuclear relaxation pathway(s) associated with the two-color signal, we simulated the non-adiabatic relaxation dynamics using trajectory surface hopping calculations within the OpenMolcas code[28,29], including the five lowest energy cationic states in all three isotopologues (see Methods for details). The active space was chosen to be large and flexible enough to describe bond breaking and formation. The initial conditions were generated using Newton-X[30], with phase space sampling performed in an uncorrelated manner based on the Wigner distribution[31]. While both $^{12}C_2H_4$ and $^{13}C_2H_4$ belong to the $D_{2h}$ symmetry point group, a single $^{13}C$-isotopic substitution leads to symmetry breaking and the resulting $^{13}C^{12}CH_4$ molecule belongs to the $C_{2v}$ point group. Due to the stochastic nature of the sampling process, the different sets of initial conditions do not strictly fulfill symmetry. To correct for this, the initial conditions of each isotopologue were symmetrized a posteriori according to its point group. In particular, the geometries explored along the non-adiabatic simulations were replicated by applying the symmetry elements (except for the identity) of the corresponding point group: seven elements (3 symmetry axes, 3 mirror planes, and the inversion) for $^{12}C_2H_4^+$ and $^{13}C_2H_4^+$, and three elements (the symmetry axis and 2 mirror planes) for $^{13}C^{12}CH_4^+$, at the equilibrium geometry (see Methods for details).

The Dyson amplitudes (an approximation of photoelectron intensities) were calculated for 100 of the initial conditions for each isotopologue from the neutral ground state to the first five electronic states of the cation. The results (see Supplementary Fig. 1) show the independence of the estimated photoelectron intensities on the isotopic substitution and the low photoionization intensity for the $D_4$ state. Only about 15% of the initial conditions for each isotopologue showed a Dyson norm greater than 0.25 for the latter cationic state, suggesting that this state contributes only moderately to the experimental signal. We calculated the time-dependent yields for all possible relaxation channels, namely, H-loss, $H_2$ and/or, 2H-loss, and ethylene–ethyldiene isomerization. In general, for relaxation from the excited cationic states, $D_1$, $D_2$, $D_3$, and $D_4$, the corresponding trends observed for three different isotopologues approximately follow the change in mass (see Supplementary Fig. 2). However, for ethylene–ethylidene conversion, the calculations show that the ethylidene population saturates faster and to a lower yield in the symmetry-broken species compared to the other two for relaxation from the $D_3$ state (see Fig. 2a–c, solid lines). Note that the ethylidene population curves increase in a non-monotonic manner; this is because the isomerization reaction is reversible. To quantitatively compare the calculations with the experiment, the theoretical ethylidene populations were fitted with a logistic function, $f(t) = \frac{P}{1+e^{-k(t-t_{1/2})}}$, where $P$ denotes the asymptotic population, $k$ is the population growth rate, and $t_{1/2}$ signifies the time to form half of the asymptotic ethylidene population. The fits reported in Fig. 2 (dashed-dotted lines) were performed assuming $P$, $k$ and $t_{1/2}$ as free parameters. To verify the robustness of the fitting procedure, we checked the fitting with normalized $P$ as well as fixed $k$ for three isotopologues. While the $R^2$ value reduces by less than 5% for constrained fitting, $t_{1/2}$ do not change significantly (see Supplementary Fig. 3a, b and Supplementary Table 1). Although $^{12}CH_3^+$ fragments from $^{13}C^{12}CH_4^+$ could not be detected due to overlapping signal from $^{13}CH_2^+$, we have calculated the ethylidene population as a function of time for the $^{13}C^{12}CH_4^+ \rightarrow {}^{12}CH_3\text{-}^{13}CH^+$ isomerization channel. The yield for this channel saturates at a similar rate compared to that for the $^{13}C^{12}CH_4^+ \rightarrow {}^{13}CH_3\text{-}^{12}CH^+$ channel (see Supplementary Fig. 4), indicating both relaxation channels to be sensitive towards the effect of isotope-induced symmetry breaking.

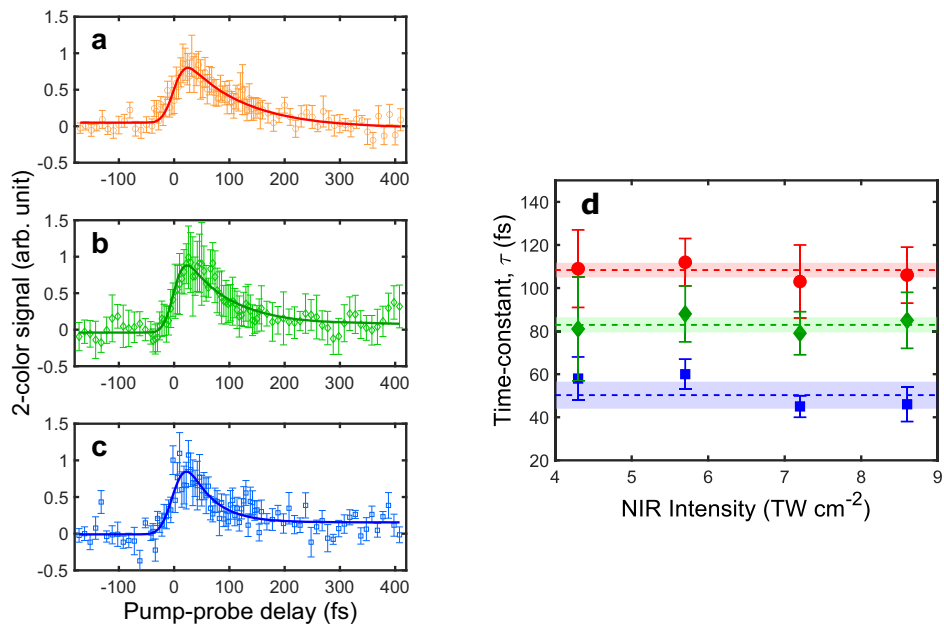

**Fig. 1 | Experimental results.** Typical two-color signal as a function of the XUV pump -- NIR probe delay for **a** $^{12}CH_3^+$ from $^{12}C_2H_4$, **b** $^{13}CH_3^+$ from $^{13}C_2H_4$ and **c** $^{13}CH_3^+$ from $^{13}C^{12}CH_4$. The error-bars correspond to statistical fluctuations. In each panel, the solid line denotes fitting with an exponentially decaying function (see Methods for details). **d** The corresponding time constants, $\tau$, at different NIR-probe intensities for three different isotopologues: $C_2H_4$ (solid red circles), $^{13}C_2H_4$ (solid green diamonds) and $^{13}C^{12}CH_4$ (solid blue squares). The uncertainties stem only from the fitting. In each case, the dashed line corresponds to the weighted average and the shaded area represents the 95% confidence interval.

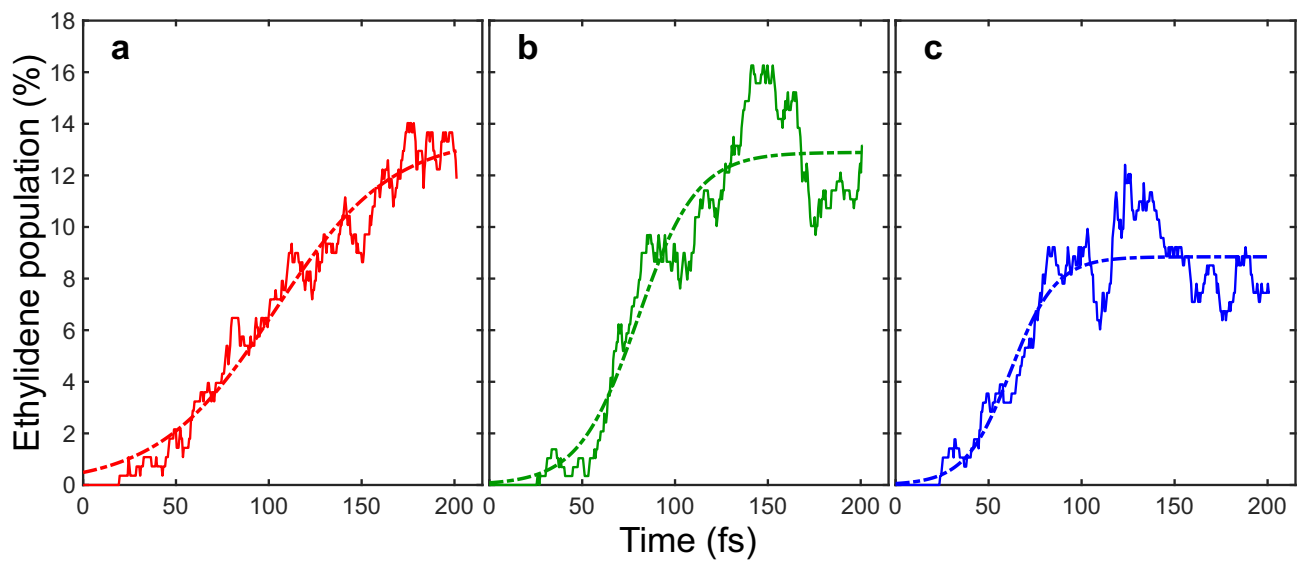

**Fig. 2 | Ethylene–ethylidene photoisomerization yield.** Theoretical ethylidene population as a function of time for the parent ions **a** $C_2H_4^+$, **b** $^{13}C_2H_4^+$, and **c** $^{13}C^{12}CH_4^+$, upon excitation to $D_3$ electronic state. The dashed-dotted lines correspond to the fitting with a logistic function (see main text for details).

The fits reflect a significantly lower asymptotic ethylidene isomer population in $^{13}C^{12}CH_4$: $P = 9$%, compared to 14% in $^{12}C_2H_4$, and 13% in $^{13}C_2H_4$. The population saturates faster: $k = 0.082$ fs$^{-1}$ in $^{13}C^{12}CH_4$, in contrast with 0.032 fs$^{-1}$ in $^{12}C_2H_4$. For $^{13}C_2H_4$, an intermediate value of 0.063 fs$^{-1}$ is obtained. More importantly, the inflection point is reached earlier in time in the symmetry-broken molecule. Indeed, $t_{1/2}$ is 62 fs in $^{13}C^{12}CH_4$, 103 fs in $^{12}C_2H_4$, and 80 fs in $^{13}C_2H_4$, which matches well experimental $\tau$-values (see, Table 1). This also implies that if the NIR-probe arrives at a delay by which the relevant excited state population in $D_3$ has relaxed via the formation of around 50% of the asymptotic ethylidene population, it can no longer efficiently produce the $CH_3^+$ ions from the ethylene cation, leading to the exponential decay observed in the experimental two-color signal. Our finding that $CH_3^+$ yield primarily reflects relaxation from a higher-lying excited cationic state, such as $D_3$, is supported by previous observation[32]. As in the measurements, while the differences in the fitted parameters between iso-symmetric $^{12}C_2H_4$ and $^{13}C_2H_4$ may be explained in terms of changes in mass, the predicted behavior of the symmetry-broken $^{13}C^{12}CH_4$ cannot be. Nevertheless, the trend observed in the experimental decay constants for three isotoplogues agree with that observed in the theoretical timescales for the ethylene-ethylidene isomerization channel. This allows us to assign the time-dependent $^{12}CH_3^+$ (or, $^{13}CH_3^+$) yield with the structural relaxation pathway via ethylene-ethylidene isomerization.

## Discussion

For ethylene and ethylene-$^{13}C_2$, the representation of the twelve normal modes in $D_{2h}$ symmetry point group is $3A_g + A_u + 2B_{1u} + B_{2g} + 2B_{2u} + 2B_{3g} + B_{3u}$. As the inversion symmetry is broken, the representation in the resulting $C_{2v}$ point group becomes $5A_1 + A_2 + 2B_1 + 4B_2$. The correlation relations are as follows: $A_g$, $B_{1u} \rightarrow A_1$; $A_u \rightarrow A_2$; $B_{2g}$, $B_{3u} \rightarrow B_1$; $B_{2u}$, $B_{3g} \rightarrow B_2$. The normal modes within symmetries $A_1$, $B_1$ and $B_2$ in $^{13}C^{12}CH_4$ are allowed, from a symmetry point of view, to couple and exchange vibrational energies with more modes compared to those having $D_{2h}$ point group, thus affecting the subsequent relaxation dynamics in the excited cation.

To investigate the effect of the breakdown of the inversion symmetry on the coupling of normal modes, we conducted a principal component analysis (PCA) on the geometries visited along the trajectories initiated upon excitation to the $D_3$ electronic state for each isotopologue (see Supplementary Note 2 for details about the PCA). The frames where H-loss and $H_2$ (or, 2H)-loss occurred were excluded from the analysis to focus only on the ethylidene conversion channel. The resulting principal components indicate the essential motions for this reaction pathway. For each isotopologue, the components were projected onto the corresponding normal modes (see Supplementary Fig. 5 for pictorial depiction of the normal modes in ethylene) and analyzed to identify which modes contribute to the ethylene–ethylidene isomerization.

The outcomes of the analysis for $D_3$ electronic state are presented in Fig. 3a, b for $^{12}C_2H_4$ and $^{13}C_2H_4$ (results for other initial electronic states are shown in Supplementary Fig. 6). These maps show the absolute values of the coefficients of each principal component (columns) in terms of normal modes (rows). The principal component (PC) analysis shows that the essential motions for the ethylidene isomerization in $^{12}C_2H_4$ are along $CH_2$ twisting normal mode (PC0), $CH_2$ scissoring, $C=C$ stretching and $CH_2$ stretching (PC1 and PC3), $CH_2$ rocking (PC2) and $CH_2$ scissoring and $CH_2$ stretching (PC4). No significant differences were observed between the two iso-symmetric isotopologues, except for permutation of some principal components with similar associated variance. However, the analysis in symmetry-broken $^{13}C^{12}CH_4$ (see Fig. 3c) displays significant mixing between normal modes (now symmetry-allowed) with the appearance of three different clusters. For example, two $CH_2$-scissoring modes with $A_g$ and $B_{1u}$ symmetries in $D_{2h}$ that transforms into $A_1$ symmetry in $C_{2v}$ now mix with each other. Looking at the principal components 1, 2 and 4 (circled region in Fig. 3c), which individually explains a variance ratio of about 0.4, 0.1, and 0.1, respectively, we found that in $^{13}C^{12}CH_4$ they contain significant contributions from five different normal modes, all of them with the $A_1$

Table 1 | Comparison between experimental and theoretical timescales

| Parent | Fragment | τ (fs) | $t_{1/2}$ (fs) | P (%) | k (fs$^{-1}$) |
|---|---|---|---|---|---|
| $^{12}C_2H_4$ | $^{12}CH_3^+$ | 108 ± 4 | 103 ± 1 | 13.5 ± 0.1 | 0.032 ± 0.001 |
| $^{13}C_2H_4$ | $^{13}CH_3^+$ | 83 ± 4 | 80 ± 1 | 12.9 ± 0.1 | 0.063 ± 0.002 |
| $^{13}C^{12}CH_4$ | $^{13}CH_3^+$ | 50 ± 8 | 62 ± 1 | 8.8 ± 0.1 | 0.082 ± 0.003 |

The experimental and theoretical timescales, along with the asymptotic population and the population growth rate. The uncertainties originate from the fitting only.

symmetry. The PCA was performed also on the dynamics upon excitation to the $D_1$, $D_2$, and $D_4$ state, for the three isotopologues, and the results are reported in Supplementary Fig. 6 and the explained variance ratio for all principal components are shown in Supplementary Fig. 7.

For the isotope-substituted molecule, the symmetry-breaking kinetic energy term in the Hamiltonian of the molecule primarily depends on the mass difference as $\propto \Delta m (\nabla_R \cdot \sum_j \nabla_j)$, where $R$ is the position of the substituted nuclei and $j$ denotes the relevant electronic coordinates[13]. For small value of the mass difference $\Delta m$, such as in the present case, it approximately preserves the overall Hamiltonian symmetry. In this situation, the pure electronic relaxation still respects the approximate symmetry; this is seen in the time-dependent electronic populations for different excited states, which do not change for the three isotopologues (Supplementary Fig. 8). However, the structural relaxation mediated by several nuclear coordinates does not respect the approximate symmetry. Hence, the isotope-induced inversion symmetry breaking is an example of explicit symmetry breaking[4], as opposed to spontaneous symmetry breaking typically associated with Jahn–Teller effects[8,10]. The XUV-induced ultrafast ethylene–ethylidene conversion process serves as an ideal proving ground for this mechanism. Using the PCA, we clearly identified normal-mode mixing as a key consequence of this explicit symmetry breaking (Fig. 3). The mixing causes an efficient intra-molecular vibrational energy redistribution[33], decreasing the overall isomerization yield and thus, the asymptotic population of ethylidene, $P$. It is worth noting that the differences between the ethylidene yields of the different isotopologues upon excitation to the $D_4$ state (see Supplementary Fig. 2) can also be rationalized considering the intramolecular vibrational energy redistribution that follows from the $D_{2h}$ to $C_{2v}$ point-group symmetry-lowering induced by a single $^{13}C$-atom substitution, suggested by the normal mode mixing in the PCs reported in Supplementary Fig. 6g–i.

To get a qualitative picture behind how the mixing between different normal modes can influence the ethylene to ethylidene conversion as reported in Fig. 2, we studied the temporal evolution of several internal (nuclear) coordinates of the photoionized molecule, upon excitation to the $D_3$ electronic state. Like in the PCA, the frames where H-loss and $H_2$ (or, 2H)-loss occurred were excluded. The three relevant internal coordinates are the absolute dihedral angle, $|\Phi|$, the average distance $d$, between carbon atom $C^1$ and hydrogen atom $H^5$, and the angle, $\alpha$ identified by atoms $H^5C^2C^1$ (see Fig. 4 insets). For all the panels in Fig. 4, the temporal variations were smoothed using a Gaussian kernel with a full width at half maximum of 36 fs (same as the XUV–NIR cross-correlation) to better reveal the underlying trend. Ethylene–ethylidene isomerization requires the combination of a rotation around $|\Phi|$, a closure of the angle $\alpha$, and a shortening of the bond length $d$. Fig. 4a shows the temporal variation of $|\Phi|$, identified by the atoms $H^3C^1C^2H^5$ and $H^3C^1C^2H^6$: the differences between the three different isotopologues are not significant. This is expected since motion around $|\Phi|$ corresponds to a normal mode of $A_u$ symmetry in $D_{2h}$, $A_2$ in $C_{2v}$, and is not allowed to couple to more modes upon breakdown of the inversion symmetry. However, for $d$ and $\alpha$ internal coordinates, clear differences between the different isotopologues can be observed (see, Fig. 4b, c). At longer times (>100 fs), the values of the coordinates $d$ and $\alpha$ decrease further in the iso-symmetric systems, in particular for $^{13}C_2H_4$. To understand these evolutions, it is important to realize that the time evolutions of average nuclear coordinates are representative of both the yields ($P$) and the rates ($k$) of the processes. The larger distortion of $d$ and $\alpha$ in the $^{13}C$ species results in a relatively large asymptotic population, $P$, and an intermediate rate, $k$. Interestingly, $^{12}C_2H_4$ and $^{13}C^{12}CH_4$ show a comparable change in the $d$ and $\alpha$ coordinates within the simulation time. This may appear strange at first glance but can actually be explained by two effects compensating each other. On one hand, the $^{12}C_2H_4$ is the isotopologue with the lowest rate, $k$, but the highest yield. On the other hand, the symmetry-broken isotopologue shows the highest rate but lowest yield (see Table 1). These two aspects of the dynamics compensate each other and lead to equivalent variations of these coordinates within the simulated time window. Now, comparing the two iso-symmetric species, they have almost the same asymptotic yield but $^{12}C_2$-isotope has a slower rate, hence the smaller distortion as a function of time. A similar trend was also found for the average of the relevant normal mode ($CH_2$ scissoring) displacements (see Fig. 4d). We performed the same analysis for all the other normal modes by filtering everything but ethylene to ethylidene conversion as well as trajectories ending in ethylidene; for both filters the effect of symmetry breaking was only observed for the relevant modes with $A_g$ symmetry, further reinforcing our interpretation (see Supplementary Figs. 9 and 10). After 100 fs, the coordinates change more slowly in the symmetry-broken system, suggesting that much of the excited-state population has already relaxed structurally by this time. For the two iso-symmetric molecules, especially $^{13}C_2H_4$ species, even at longer times, there is still a significant excited-state population left to decay, causing noticeable changes in the coordinates. As pointed out earlier, this is reflected in the experimentally measured faster decay timescale of the $CH_3^+$ ion yield ($\tau$) with respect to the symmetry-intact molecules.

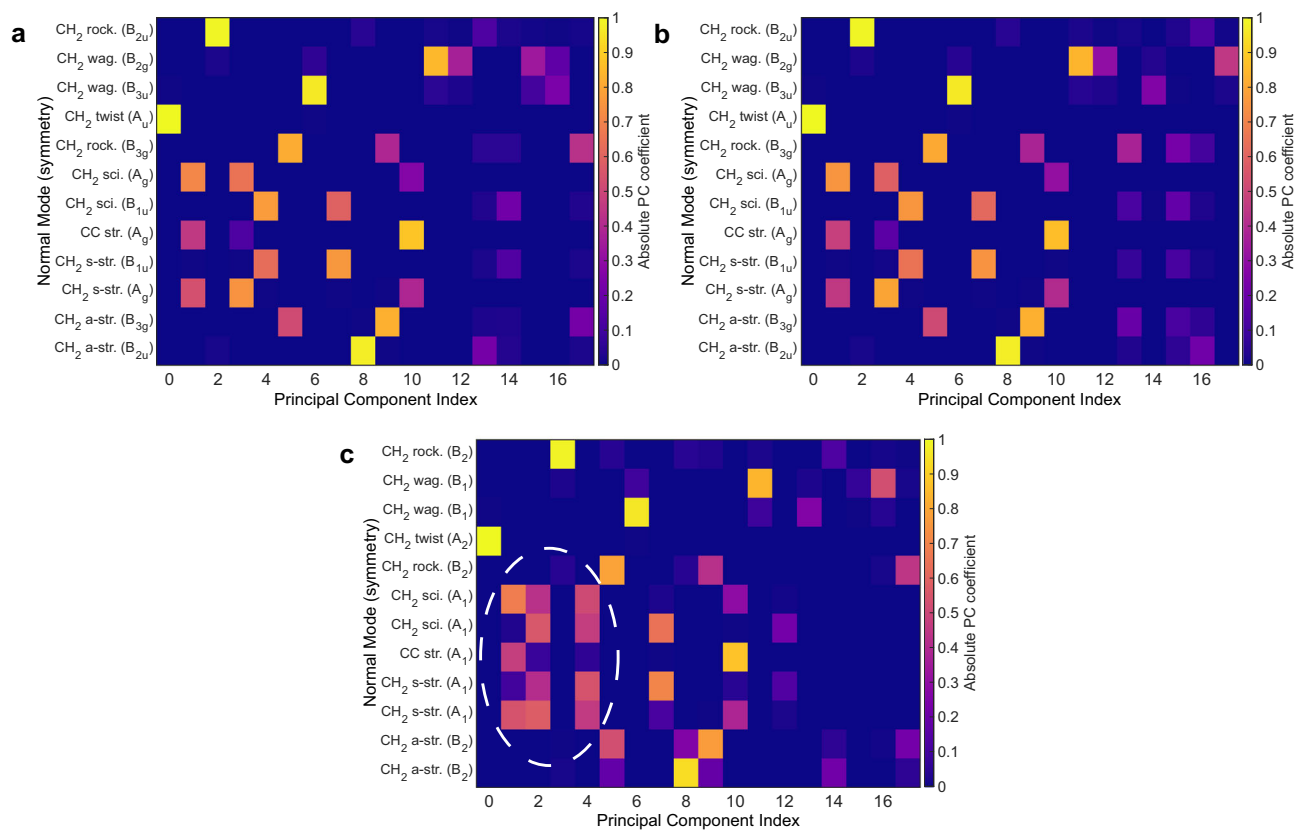

**Fig. 3 | Principal component analysis.** The absolute values of the principal component coefficients with $D_3$ initial electronic state for **a** $C_2H_4$, **b** $^{13}C_2H_4$, and **c**, $^{13}C^{12}CH_4$. Note the mixing between different normal modes (denoted by dashed circle) in ethylene-$^{13}C$ following the loss of the g/u-symmetry. The explained variance ratio for different principal components are shown in Supplementary Fig. 7.

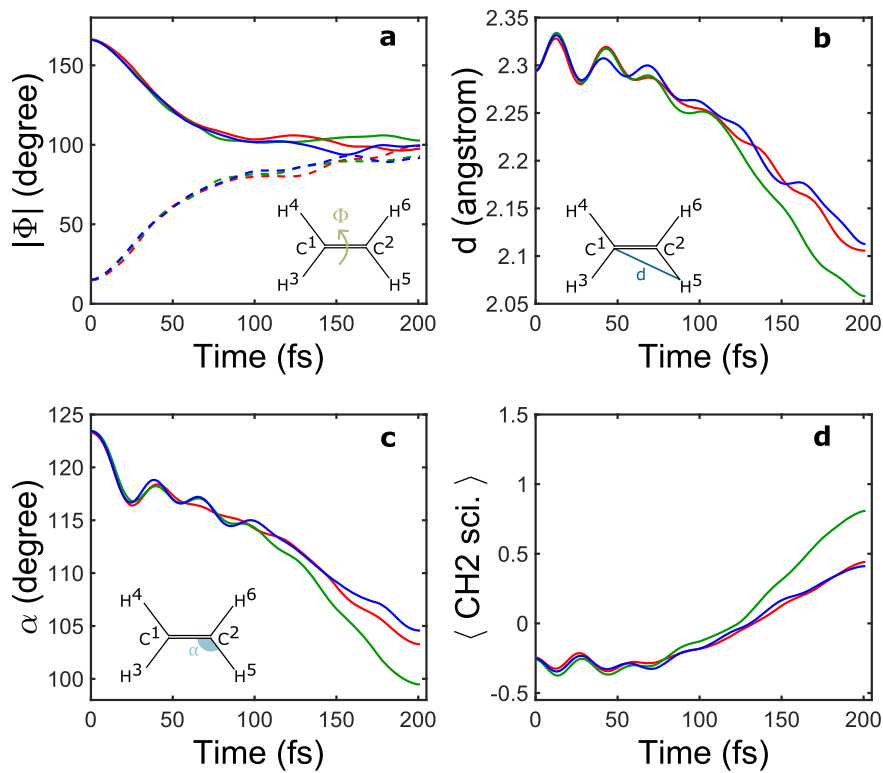

**Fig. 4 | Temporal variation of nuclear coordinates. a** The temporal evolution of the dihedral angle, $|\Phi|$, defined by $H^3C^1C^2H^6$ (solid lines) and $H^3C^1C^2H^5$ (dashed lines) in different isotopologues ($^{12}C_2H_4$ in red, $^{13}C_2H_4$ in green, and $^{13}C^{12}CH_4$ in blue) upon excitation to the $D_3$ electronic state. **b** The same as **a** but for the distance, $d$, between carbon atom $C^1$ and hydrogen atom $H^5$. **c** Once again the same as **a** but for the angle, $\alpha$, identified by atoms $H^5C^2C^1$. The inset in each case shows the corresponding definition for the coordinate. **d** The time dependent variation of the average normal mode ($CH_2$ scissoring) displacements. In all the analyses the frames in which H-loss and $H_2$ or, 2H-loss takes place are excluded.

We have shown that isotope-induced explicit symmetry breaking can significantly alter structural relaxation timescales, despite minor changes in the overall mass, due to mixing of normal modes in the photo-excited molecular cation. While the decay of electronic populations remains unaffected by the mixing, it can significantly suppress photo-isomerization yields via intra-molecular vibrational energy redistribution. Isotope substitution has previously been used to change the outcome of light-induced electron transfer processes in molecules[34]. Our results highlight the potential of symmetry breaking to control similar phenomena in molecules. With the advent of brighter light sources having shorter wavelengths and ultrashort pulse duration, the present study can act as a benchmark not only for future experiments on isotope-induced symmetry-breaking in more complex systems but also to design new photoresponsive molecules based on specific symmetries.

## Methods
### Experimental setup
The experiment was performed at the Circé attosecond beamline at the Institut Lumière Matière, Lyon[35]. The output of a commercially available Ti:Saph amplifier laser system (Coherent Legend), providing NIR pulses (wavelength: centered around 800 nm) at a repetition rate of 5 kHz with 2 mJ energy per pulse, was divided into two parts along the arms of a Mach-Zehnder interferometer. In the pump arm, the XUV attosecond pulse train was generated by focusing the NIR beam unto a jet of noble gas atoms (either xenon or, krypton). The attosecond pulse trains produced via HHG represent a comb of odd-order high harmonics in the frequency domain, out of which only three, namely 11, 13 and 15 were selected using a tin filter (see Supplementary Fig. 11 for the HHG spectra). The Sn-filter also removes the co-propagating NIR-pulse. Using the chirp-scan technique[36], we determined the temporal duration of the NIR-pulse to be 27.5 ± 0.2 fs (see Supplementary Fig. 12 for the NIR temporal profile). In the probe arm, the remaining NIR beam is focused with a 1-m lens onto the interaction region of a velocity map imaging (VMI) spectrometer. The ionic fragments were collected across $4\pi$ solid angle by the VMI spectrometer operating in the time-of-flight mode. We procured the $^{13}C$-isotope substituted samples from Sigma-Aldrich (CAS 6145-18-2 and 51915-19-6) and used them without further purification. The molecules were introduced in the interaction region as an effusive jet via a 200 μm hole in the repeller plate of the VMI spectrometer. The values of the probe intensity reported in Fig. 1d were estimated to have an uncertainty up to 20%. The cross-correlation between the XUV pump and NIR probe was 35 ± 1 fs (see Supplementary Fig. 13 for cross-correlation measurements in argon).

### Data analysis

The data reported in Fig. 1a–c consists of an average over 10 consecutive sets of forward ($-160$ fs to $+400$ fs) and backward scans ($+400$ fs to $-160$ fs), thus, in total 20 scans across the entire range of the pump-probe delay, at each probe intensity. The error-bars corresponds to the statistical fluctuations for these 10 sets of scan, obtained as the standard error. Pure two-color signal was obtained by subtracting the XUV-only and NIR-only yield from the time dependent signal at each pump-probe delay. The XUV-only time-of-flight mass spectra normalized on the parent-ion-yield for all three isotopologues are shown in Supplementary Fig. 14. The two-color signal as a function of the pump-probe delay was fitted with an exponential decay function convoluted (*) with a normalized Gaussian denoting the instrument response function, $g_{IRF}(t) = \frac{1}{\sigma_p \sqrt{2\pi}} \exp\left[-\left(\frac{t}{\sigma_p \sqrt{2}}\right)^2\right]$. The fitting function can, therefore, be written as, $g(t) = \left[\{A \exp\left(-\frac{t-t_0}{\tau}\right) + B\}\Theta(t-t_0)\right] * g_{IRF}(t) + C$. Here, $\tau$ is the time-constant, $2.355\sigma_p$ provides a measure of the cross-correlation, the temporal overlap between the pump and the probe is given by $t_0$, and $\Theta(t)$ denotes a Heaviside step function. See Supplementary Note 3 for details about the precise determination of $t_0$ and $\sigma_p$. The linear background, C, was subtracted from the two-color signal for each measurement. The error-bar in $\tau$ for an individual measurement shown in Fig. 1d represents uncertainties obtained from the fitting procedure using the least squares method. The dashed line is the average, weighted against these uncertainties. The error on the weighted average was determined by the corresponding standard deviations of the four independent measurements.

### Theoretical methods

We performed non-adiabatic dynamics simulations of all the $^{13}C$-containing isotopologues of ethylene cation examined in this study ($^{12}C_2H_4^+$, $^{13}C_2H_4^+$, and $^{13}C^{12}CH_4^+$). The simulations were conducted using the surface hopping method, with the Tully fewest switches algorithm[37], within the OpenMolcas[28,29] software (version 22.10-354-g7f2c128[38]). The trajectories were integrated with a nuclear time step of 20 a.u. for ca. 203 fs. Each nuclear time step was divided into 96 substeps for the electronic wave function integration, carried out using the Hammes-Schiffer Tully[39] scheme with biorthonormalization, as implemented in OpenMolcas[40]. To account for decoherence effects, the Granucci and Persico energy-based correction[41] was applied with a decay factor of 0.1 Hartree. After a hop, the entire classical velocity vector is rescaled to ensure energy conservation along the trajectory, and frustrated hops are ignored. The electronic structure of all the studied isotopologues was treated at the Complete Active Space SCF[42] (CASSCF) level, with state-averaging performed over the five lowest-energy cationic states. See Supplementary Table 2 for the theoretical (vertical) ionization energies of ethylene from its neutral ground state to the first five cationic states. The selection of the active space, consisting of 11 electrons in 12 orbitals (the $\sigma$ and $\sigma^*$ orbitals of the four C–H bonds, and the $\sigma$, $\sigma^*$, $\pi$, and $\pi^*$ orbitals of the C = C double bond), was based on previous works by some of the present authors[17,20]. Indeed, the employed active space was demonstrated to be flexible enough to describe bond breaking and formation. While dynamic correlation may affect the XUV-induced dynamics[43], previous works investigating the influence of the dynamical correlation introduced at XMS-CASPT2 level[20] showed only a modest reduction of the $D_0$-$D_1$ energy gap at Franck-Condon geometry. Moreover, a semi-quantitative agreement between CASSCF and CASPT2 potential energy scans were found along the different relaxation pathways including the ethylene-ethylidene isomerization. The calculations employed the ANO-RCC-VDZP basis set[44] and Cholesky decomposition[45] to reduce computational cost.

The frequencies used for this sampling were calculated at the ground-state equilibrium geometries of the respective neutral isotopologues. For the $^{13}C_2H_4^+$ and $^{13}C^{12}CH_4^+$ isotopologues, 300 pairs of coordinates and velocities were generated and propagated starting from the $D_1$ and $D_3$ electronic states, 200 trajectories were initiated on $D_0$ and $D_2$. For the $^{12}C_2H_4^+$ isotopologue, the ensemble of trajectories obtained in a previous work[20], which followed the same computational protocol, was used and analyzed. The set includes 300 trajectories started on the $D_1$, $D_2$, and $D_3$ electronic states, and 200 initiated on the $D_0$ state. Finally, 120 additional trajectories were initiated on the $D_4$ electronic state, even if this state showed very low photo-ionization probability (estimated as Dyson intensities on 100 of the initial conditions for each isotopologue) compared to the other three states (see Supplementary Fig. 1). A small number of trajectories, specifically 0.0%, 0.3%, 3.3%, 7.3%, and 10.9 % of those initiated on $D_0$, $D_1$, $D_2$, $D_3$, and $D_4$, respectively, for the $^{12}C_2H_4^+$ isotopologue, exhibited a change in total energy exceeding 0.5 eV during the dynamics. Similar percentages were observed for the other two isotopologues, and these trajectories were discarded in the analysis process.

To gain insights into the reactive events in the collected ensemble of the trajectories, accordingly to a previous work[20], the ones with exactly three hydrogen atoms within 2.9 Å (value corresponding to the sum of the van der Waals radii of H and C atoms) from their nearest carbon atom were considered as having undergone H-loss. Similarly, trajectories with exactly two hydrogen atoms within 2.9 Å of their nearest carbon atom were considered to undergo through $H_2$-loss or 2H-loss. Finally, trajectories where all four hydrogen atoms are less than 2.9 Å from their nearest carbon atom, with three hydrogen atoms being closer to one carbon than the other, were classified as the ethylidene form. To account for the symmetries in the isotopologues, the geometries explored along the non-adiabatic simulations were replicated by applying the symmetry elements of the corresponding point group at the equilibrium geometry. For each explored geometry of $^{12}C_2H_4^+$ and $^{13}C_2H_4^+$, seven additional geometries were generated by performing the symmetry operations corresponding to the three symmetry axes, three mirror planes, and the inversion. In case of $^{13}C^{12}CH_4^+$, only three additional geometries were produced by performing the symmetry operations corresponding to the symmetry axis and two mirror planes. This enabled us to account for the nuclear symmetry of the investigated isotopologues. The statistical analyses were performed keeping the ensembles of trajectories that started upon excitation on $D_0$, $D_1$, $D_2$, $D_3$, and $D_4$ separated for each isotopologue. In the case of the principal component analysis, conducted on each ensemble separately, the frames of the corresponding trajectories were joined, discarding the ones in which H-loss or 2H and/or $H_2$-loss took place.

To estimate the photoelectron intensities, we calculated the Dyson amplitudes for 100 of the initial conditions for each isotopologue from the neutral ground state to the first five electronic states of the cation at the same level of theory used for the non-adiabatic dynamics simulations, i.e., SA5-CASSCF(11e,12o). The geometries obtained from the symmetrization of the initial conditions exhibit the same Dyson norm of the corresponding Wigner-sampled structure. Finally, initial conditions were not selected on the basis of its Dyson norm.

## Data availability
Any additional data related to the findings are available from the corresponding authors upon request.

## Code availability
Codes developed for this study are available from the corresponding authors upon request.

## References

1. Currie, P. Sur la symétrie dans les phénomènes physiques, symétrie d'un champ électrique et d'un champ magnétique. *J. Phys. Théo. Appliquée* **3**, 393 (1894).
2. Bardeen, J., Cooper, L. N. & Schrieffer, J. R. Microscopic theory of superconductivity. *Phys. Rev.* **106**, 162 (1957).
3. Higgs, P. W. Broken symmetries and the masses of gauge bosons. *Phys. Rev. Lett.* **13**, 508 (1964).
4. Gross, D. J. The role of symmetry in fundamental physics. *Proc. Natl. Acad. Sci. USA* **93**, 14256 (1996).
5. Martín, F. et al. Single photon-induced symmetry breaking of $H_2$ dissociation. *Science* **315**, 629 (2007).
6. González-Castrillo, A., Palacios, A., Bachau, H. & Martín, F. Clocking ultrafast wave packet dynamics in molecules through UV-induced symmetry breaking. *Phys. Rev. Lett.* **108**, 063009 (2012).
7. Palacios, A., Rivière, P., González-Castrillo, A. & Martín, F. XUV lasers for ultrafast electronic control in $H_2$. in *Ultrafast Phenomena in Molecular Sciences* Vol. 107 (eds de Nalda, R. & Bañares, L.) (Springer, 2014).
8. Ridente, E. et al. Femtosecond symmetry breaking and coherent relaxation of methane cations via X-ray spectroscopy. *Science* **380**, 713 (2023).
9. Pertot, Y. et al. Time-resolved X-ray absorption spectroscopy with a water window high-harmonic source. *Science* **355**, 264 (2017).
10. Li, M. et al. Ultrafast imaging of spontaneous symmetry breaking in a photoionized molecular system. *Nat. Comm.* **12**, 4233 (2021).
11. Haase, D., Manz, J. & Tremblay, J. C. Attosecond charge migration can break electron symmetry while conserving nuclear symmetry. *J. Phys. Chem. A* **124**, 3329 (2020).
12. Livshits, E. et al. Symmetry-breaking dynamics of a photoionized carbon dioxide dimer. *Nat. Comm.* **15**, 6322 (2024).
13. Bunker, P. R. Forbidden transitions in homopolar isotopically unsymmetric diatomic molecules and the dipole moment of HD. *J. Mol. Spectrosc.* **46**, 119 (1973).
14. Reinhold, E., Hogervorst, W. & Ubachs, W. Complete g–u symmetry breaking in highly excited valence states of the HD molecule. *Chem. Phys. Lett.* **296**, 411 (1998).
15. Hölsch, N. et al. Ionization and dissociation energies of HD and dipole-induced g/u-symmetry breaking. *Phys. Rev. A* **108**, 022811 (2023).
16. Rolles, D. et al. Isotope-induced partial localization of core electrons in the homonuclear molecule $N_2$. *Nature* **437**, 711–715 (2005).
17. Vacher, M., Boyer, A., Loriot, V., Lépine, F. & Nandi, S. Few-femtosecond isotope effect in polyatomic molecules ionized by extreme ultraviolet attosecond pulse trains. *J. Phys. Chem. A* **126**, 5692 (2022).
18. Gong, X. et al. Attosecond delays between dissociative and non-dissociative ionization of polyatomic molecules. *Nat. Comm.* **14**, 4402 (2023).
19. Ertel, D. et al. Influence of nuclear dynamics on molecular attosecond photoelectron interferometry. *Sci. Adv.* **9**, adh7747 (2023).
20. Fransén, L., Tran, T., Nandi, S. & Vacher, M. Dissociation and isomerization following ionization of Ethylene: insights from nonadiabatic dynamics simulations. *J. Phys. Chem. A* **128**, 1457 (2024).
21. Ludwig, A. et al. Ultrafast relaxation dynamics of the ethylene cation $C_2H_4^+$. *J. Phys. Chem. Lett.* **7**, 1901 (2016).
22. van Tilborg, J. et al. Femtosecond isomerization dynamics in the ethylene cation measured in an EUV-pump NIR-probe configuration. *J. Phys. B At. Mol. Opt. Phys.* **42**, 081002 (2009).
23. Ibuki, T., Cooper, G. & Brion, C. E. Absolute dipole oscillator strengths for photoabsorption and the molecular and dissociative photoionization of ethylene. *Chem. Phys.* **129**, 295 (1989).
24. Hervé, M. et al. Ultrafast dynamics of correlation bands following XUV molecular photoionization. *Nat. Phys.* **17**, 327 (2021).
25. Murillo-Sánchez, M. L. et al. Femtosecond XUV-IR induced photodynamics in the methyl iodide cation. *New J. Phys.* **23**, 073023 (2021).
26. Vismarra, F. et al. Few-femtosecond electron transfer dynamics in photoionized donor-$\pi$-acceptor molecules. *Nat. Chem.* **16**, 2017 (2024).
27. Felker, P. M. & Zewail, A. H. Rates of photoisomerization of trans-stilbene in isolated and solvated molecules: experiments on the deuterium isotope effect and RRKM behavior. *J. Phys. Chem.* **89**, 5402 (1985).
28. Fdez Galván, I. et al. OpenMolcas: from source code to insight. *J. Chem. Theo. Comput.* **15**, 5925 (2019).
29. Li Manni, G. et al. The OpenMolcas web: a community-driven approach to advancing computational chemistry. *J. Chem. Theory Comput.* **19**, 6933 (2023).
30. Barbatti, M. et al. Newton-X: a surface-hopping program for nonadiabatic molecular dynamics. *Wiley Interdiscip. Rev. Comput. Mol. Sci.* **4**, 26 (2014).
31. Wigner, E. On the quantum correction for thermodynamic equilibrium. *Phys. Rev.* **40**, 749 (1932).
32. Lucchini, M. et al. Few-femtosecond $C_2H_4^+$ internal relaxation dynamics accessed by selective excitation. *J. Phys. Chem. Lett.* **13**, 48 (2022).
33. Boyer, A. et al. Ultrafast vibrational relaxation dynamics in XUV-excited polycyclic aromatic hydrocarbon molecules. *Phys. Rev. X* **11**, 041012 (2021).
34. Delor, M. et al. Directing the path of light-induced electron transfer at a molecular fork using vibrational excitation. *Nat. Chem.* **9**, 1099 (2017).
35. Loriot, V. et al. Attosecond metrology of the two-dimensional charge distribution in molecules. *Nat. Phys.* **20**, 765 (2024).
36. Loriot, V., Gitzinger, G. & Forget, N. Self-referenced characterization of femtosecond laser pulses by chirp scan. *Opt. Exp.* **21**, 24879 (2013).
37. Tully, J. C. Molecular dynamics with electronic transitions. *J. Chem. Phys.* **93**, 1061 (1990).
38. OpenMolcas v22.10. https://gitlab.com/Molcas/OpenMolcas/-/releases/v22.1web0 (October, 2022).
39. Hammes-Schiffer, S. & Tully, J. C. Proton transfer in solution: molecular dynamics with quantum transitions. *J. Chem. Phys.* **101**, 4657 (1994).
40. Merritt, I. C. D., Jacquemin, D. & Vacher, M. Nonadiabatic coupling in trajectory surface hopping: how approximations impact excited-state reaction dynamics. *J. Chem. Theory Comput.* **19**, 1827 (2023).
41. Granucci, G. & Persico, M. Critical appraisal of the fewest switches algorithm for surface hopping. *J. Chem. Phys.* **126**, 134114 (2007).
42. Roos, B. O., Taylor, P. R. & Sigbahn, P. E. A Complete Active Space SCF Method (CASSCF) using a density matrix formulated super-CI approach. *Chem. Phys.* **48**, 157 (1980).
43. Ibele, L. M., Memhood, A., Levine, B. G. & Avagliano, D. Ab initio multiple spawning nonadiabatic dynamics with different CASPT2 flavors: a fully open-source PySpawn/OpenMolcas interface. *J. Chem. Theo. Comput.* **20**, 8140 (2024).
44. Roos, B. O., Lindh, R., Malmqvist, P.-Å., Veryazov, V. & Widmark, P.-O. Main group atoms and dimers studied with a new relativistic ANO basis set. *J. Phys. Chem. A* **108**, 2851 (2004).
45. Aquilante, F., Gagliardi, L., Pedersen, T. B. & Lindh, R. Atomic cholesky decompositions: a route to unbiased auxiliary basis sets for density fitting approximation with tunable accuracy and efficiency. *J. Chem. Phys.* **130**, 154107 (2009).

## Acknowledgements
We thank Emilien Prost for assistance during experiments. S.N. thanks Center National de la Recherche Scientifique (CNRS) and Fédération de Recherche André Marie Ampère, Lyon for financial support. We acknowledge support from Agence National de la Recherche (ANR-20-CE29-0021, ANR-20-CE29-0021 and ANR-21-CE30-0052). The simulations in this work were performed using HPC resources from CCIPL (Le center de calcul intensif des Pays de la Loire) and from GENCI-IDRIS (grant 2021-101353). The project is partly funded by the European Union (ERC, 101040356, ATTOP, M.V. and A.N.N.). Views and opinions expressed are those of the authors only and do not necessarily reflect those of the European Union or the European Research Council Executive Agency. Neither the European Union nor the granting authority can be held responsible for them. A.N.N. and M.V. wish to thank Anthony Ferté and Lina Fransén for stimulating discussions.

## Author contributions
A.B., Y.H., V.L., F.L., and S.N. carried out the experiment. S.N. analyzed the experimental data. The theoretical calculations were performed by A.N.N. under the supervision of M.V. A.N.N., A.B., F.L., M.V., and S.N. interpreted the results. A.N.N., M.V., and S.N. wrote the manuscript, which all authors discussed. S.N. conceived and led the project.

## Competing interests
The authors declare no competing interests.

## Additional information
**Supplementary information** The online version contains supplementary material available at https://doi.org/10.1038/s42004-025-01621-z.

**Correspondence** and requests for materials should be addressed to Morgane Vacher or Saikat Nandi.

**Peer review information** *Communications Chemistry* thanks Patricia Vindel-Zandbergen, Amit Samanta and the other, anonymous, reviewer(s) for their contribution to the peer review of this work. A peer review file is available.

**Reprints and permissions information** is available at http://www.nature.com/reprints

**Publisher's note** Springer Nature remains neutral with regard to jurisdictional claims in published maps and institutional affiliations.